\documentclass[epsf,twocolumn,showpacs,preprintnumbers,prb]{revtex4-1}
\usepackage{graphicx}
\usepackage{dcolumn}
\usepackage{color}

\begin{document}

\title{Symmetry preserving lattice collapse in tetragonal SrFe$_{2-x}$Ru$_x$As$_2$ ($x$ = 0, 0.2) -- \\
a combined experimental and theoretical study}

\author{Deepa Kasinathan$^1$}
\email{Deepa.Kasinathan@cpfs.mpg.de}
\author{Miriam Schmitt$^1$}
\author{Klaus Koepernik$^2$}
\author{Alim Ormeci$^1$}
\author{Katrin Meier$^1$}
\author{Ulrich Schwarz$^1$}
\author{Michael Hanfland$^3$}
\author{Christoph Geibel$^1$}
\author{Yuri Grin$^1$}
\author{Andreas Leithe-Jasper$^1$}
\author{Helge Rosner$^1$}
\email{Helge.Rosner@cpfs.mpg.de}
\affiliation{$^1$ Max Planck Institute for Chemical Physics of Solids, N\"othnitzer Str. 40, 01187 Dresden, Germany \\
$^2$ IFW Dresden, P.O. Box 270116, 01171 Dresden, Germany \\
$^3$ ESRF, BP 220, 38043 Grenoble Cedex 9, France}
\date{\today}

\begin{abstract}
In a joint experimental and theoretical study, we investigate the
isostructural collapse from the ambient pressure tetragonal
phase to a collapsed tetragonal phase for non-superconducting metallic 
SrFe$_2$As$_2$ and
SrFe$_{1.8}$Ru$_{0.2}$As$_2$. The crystallographic details
have been studied using X-ray powder diffraction up to 20 GPa
pressure in a diamond anvil cell. The structural phase transition occurs
at 10 GPa and 9 GPa for SrFe$_2$As$_2$ and
SrFe$_{1.8}$Ru$_{0.2}$As$_2$, respectively. The changes in the
unit cell dimensions are highly anisotropic with a continuous decrease 
of the $c$ lattice parameter with pressure, while the $a$-axis length increases 
until the transition to a collapsed tetragonal phase and then continues to decrease. 
Across the phase transition, we observe a volume reduction of 5\% and 4\% for
SrFe$_2$As$_2$ and
SrFe$_{1.8}$Ru$_{0.2}$As$_2$, respectively.
We are able to
discern that Ru substitution on the Fe-site acts like `chemical
pressure' to the system. 
Density-functional theory-based calculations of the electronic structure 
and electron localizability indicator are consistent with the experimental
observations.
 Detailed analysis of the electronic
structure in $k$-space and real space reveals As
4$p_{z}$ interlayer bond formation as the driving force of the
$c/a$ collapse with a change in the As-As bond length of about 0.35\AA.

\end{abstract}

\maketitle
\section{Introduction}

The surprising discovery of superconductivity in Fe-based
pnictides and chalcogenides has steered and revived the interest
in high-temperature superconductivity.\cite{Kamihara} Most of
the superconducting members discovered in the past 3 years can
be grouped into five families named by their parent compounds: 
(i) 1111 family -- $R$OFeAs/$A$FeAsF
($R$ = rare-earth, $A$=alkaline-earth metal);  (ii) 122 family -- $A$Fe$_2$As$_2$
($A$ = alkaline-earth or divalent rare-earth metal); (iii) 111 family -- $A$FeAs
($A$ = alkali-metal); (iv) 11 family -- FeSe(Te); (v) P22 family -- $P$Fe$_2$As$_2$ 
($P$ = perovskite-oxide like Sr$_{4}$Sc$_2$O$_{6}$).   The
basic common feature of these new parent compounds is
the FeAs building block separated by spacer layers comprising of
the above mentioned alkali, alkaline-earth, rare-earth
oxide/fluoride or a perovskite-oxide.\cite{Kamihara, Rotter2008,
Jeevan, FeSe, Ogino} Superconductivity is obtained by suitable doping of the
parent compounds. For selected recent exhaustive overviews, we
refer the readers to Refs.\,\onlinecite{Johnston2010, Mizuguchi2010}. The
thickness of the spacer layers governs the extent of the
quasi-two dimensional (2D) nature of the electronic structure.
Using plasma frequencies as a tool to identify the `effective
dimensionality' among these systems, it has been shown that the
1111 systems are considerably more 2D than the 122 systems.\cite{NJP}
Interestingly, within these two families a correlation between
the dimensionality and the superconducting transition
temperature ($T_{\rm c}$) can be construed, with the more 2D-like 
1111 systems exhibiting a larger $T_{\rm c}$ than the
less anisotropic 122 systems.\cite{NJP} Superconductivity in these
systems upon the suppression of spin-density-wave (SDW) antiferromagnetic order of
Fe can be realized via hole-doping, electron-doping,
isovalent substitution, or
pressure.\cite{Jeevan2,Rotter2008,Sefat2008,SrFeCoAs,SrFeRuAs,Sharma2010,pressure}
In contrast to doping, high pressure techniques provide a
cleaner route to modify the electronic structure without the
added effects of chemical complexity.\cite{Chu2009}  Many of the
high-pressure experiments were focused on investigating the
suppression of the Fe-SDW order and the enhancement of $T_{\rm
c}$. Recently, pressure-induced isostructural transitions from a
paramagnetic tetragonal phase (T) to a collapsed
tetragonal (cT) phase have been reported for several members of the
122 family: CaFe$_2$As$_2$, BaFe$_2$As$_2$,
EuFe$_2$As$_2$, and EuCo$_2$As$_2$
(Refs.\,\onlinecite{ca-pressure, ba-pressure1, ba-pressure2,
eu-pressure, co-pressure, Goldman2009}). The high electronic
flexibility of compounds with the ThCr$_2$Si$_2$-type crystal structure can
be related to the subtle interplay of covalent, ionic and
metallic bonding contributions as has been outlined in several
previous studies.\cite{Hoffmann1985, Gustenau1997a,
Gustenau1997b, Johrendt1997,Singh2008a, Singh2009a, Subedi2008,
An2009} Hoffmann and Zheng\cite{Hoffmann1985}  also pointed
out that in certain cases this might lead to structural phase
transitions (see below).  Indeed, these structural phase
transitions of first and second order have been observed in many $AT_2$P$_2$
compounds\cite{Huhnt1998} between two crystallographically isostructural
modifications with significantly different P-P distances along the
tetragonal $c$ axis.  Common to them are the dramatic and
highly anisotropic changes in unit cell dimensions (changes of
the tetragonal $c$ parameter by -10\% and counteracting changes
of $a$ by +2\%, leading to a collapse of the unit cell volume)
depending on temperature,\cite{Huhnt1997}
pressure,\cite{Huhnt1997b,Huhnt1998,Chefki1998,Ni2001}
composition,\cite{Wurth1997} and chemical
pressure.\cite{Jia2009} Curiously, the order of the
phase transition (first or second-order) was suggested to be
dependent on the transition metal atom (LaFe$_2$P$_2$ -
first-order, LaCo$_2$P$_2$ - second-order).\cite{Huhnt1998}
In contrast, recent experiments report the transition to be of
second order for the arsenides Eu$T_2$As$_2$ ($T$ = Fe, Co).
More systematic studies are necessary to clarify the nature of
these transitions, as well as to understand the underlying
physics that accompany the formation of the cT-phase. 

In our
joint theoretical and experimental study on SrFe$_2$As$_2$
and the isovalent Ru substituted sample
SrFe$_{1.8}$Ru$_{0.2}$As$_2$, we attempt to address a
multitude of issues. First we analyze the possibility of a
T $\longrightarrow$ cT phase transition in these systems as has
been suggested for many other compounds with ThCr$_2$Si$_2$-type 
crystal structure. Previous theoretical studies\cite{NJP,Valenti} have predicted a phase
transition from the magnetically-ordered orthorhombic SDW phase
to a tetragonal phase, but theoretical studies on
T $\longrightarrow$ cT phase are rather cursory. Secondly, we
address the nature of the chemical bonding in the 122-arsenides and
compare it to the well studied 122-phosphides. Finally, we challenge
the concept of chemical pressure by isovalent substitution of Fe
by Ru and study the nature of the isostructural phase transition
and the bonding situation of the pnictide. For an isovalent substitution on the Fe site, 
the As layers 
remain intact and unaffected from impurities arising from the substitution
elements.

\section{Methods}

Polycrystalline samples have been synthesized using 
solid-state reactions, similar to those described in
Refs.\,\onlinecite{NJP,SrFeRuAs}. Samples were obtained in the
form of sintered pellets. X-ray diffraction measurements (XRD)
were performed at the high-pressure beam-line ID09 of the ESRF
up to 20\,GPa at room temperature for SrFe$_2$As$_2$ and
SrFe$_{1.8}$Ru$_{0.2}$As$_2$. For best possible hydrostatic
conditions we used a membrane diamond anvil cell (DAC) with
helium as  pressure transmitting medium. The pressure was
determined using the ruby fluorescence method. The measured
powder rings were integrated using the program {\small
FIT}2{\small D}.\cite{fit2d} After a background correction the
lattice parameters were determined with the
FullProf\cite{fullprof} package.

Density-functional theory (DFT) based band-structure
calculations were performed using a full-potential all-electron
local-orbital code {\small FPLO}\cite{fplo1,
fplo2} within both local density approximation (LDA) and as
well as generalized gradient approximation (GGA). Relativistic
effects were incorporated on a scalar-relativistic level. A well
converged $k$-mesh with 24$^{3}$ points in the 
full Brillouin zone was used. The crystal structures were
optimized at different levels to investigate or isolate effects
that may depend sensitively to certain structural features. The
full-relaxation of the unit cell involves optimizing the $c/a$
ratios in addition to relaxing the $z$(As) coordinate.
The electron localizability indicator/function (ELI/ELF)
was evaluated according to
Ref.~\onlinecite{Kohout04} with an ELI/ELF module implemented within
the FPLO program package.\cite{Ormeci06} The topology of ELI was
analyzed using the program Basin\,\cite{Kohout08} with consecutive
integration of the electron density in basins, which are bound by
zero-flux surfaces in the ELI gradient field. This procedure, similar
to the one proposed by Bader for the electron density\,\cite{Bader90}
allows to assign an electron count for each basin.

\section{Results and Discussion}
\subsection{Symmetry preserving lattice collapse}
\subsubsection{\rm Experiment: SrFe$_{2}$As$_{2}$}


\begin{figure}[t]
\begin{center}
\includegraphics[%
  clip,
  width=9cm,
  angle=-0]{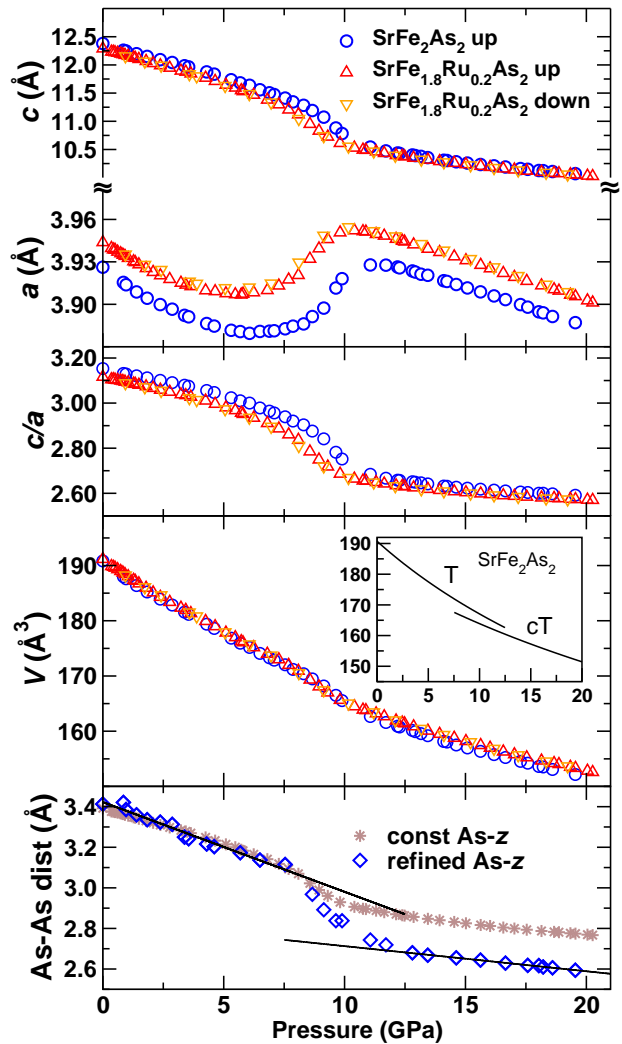}
\end{center}
\caption{\label{exp}(Color Online) The measured tetragonal lattice parameters
{\it a} and {\it c}, the axial ratio {\it c/a} and the unit cell volume as a
function of applied pressure 
for SrFe$_2$As$_2$ and
 SrFe$_{1.8}$Ru$_{0.2}$As$_2$. The measurements were
performed at room temperature. The inset shows the Birch-Murnaghan equation of
state fitting for the tetragonal (T) and collapsed (cT) phase for SrFe$_2$As$_2$.
The lowest panel displays As-As distance for SrFe$_2$As$_2$ obtained from
a full refinement of the $z$ coordinate of the As atoms. To separate the influence due to the decrease in $c$, only, the As-As
distance using the $z$ coordinate of As at ambient pressure is also shown.  }
\end{figure}

\begin{figure}[t]
\begin{center}
\includegraphics[%
  clip,
  width=9cm,
  angle=-0]{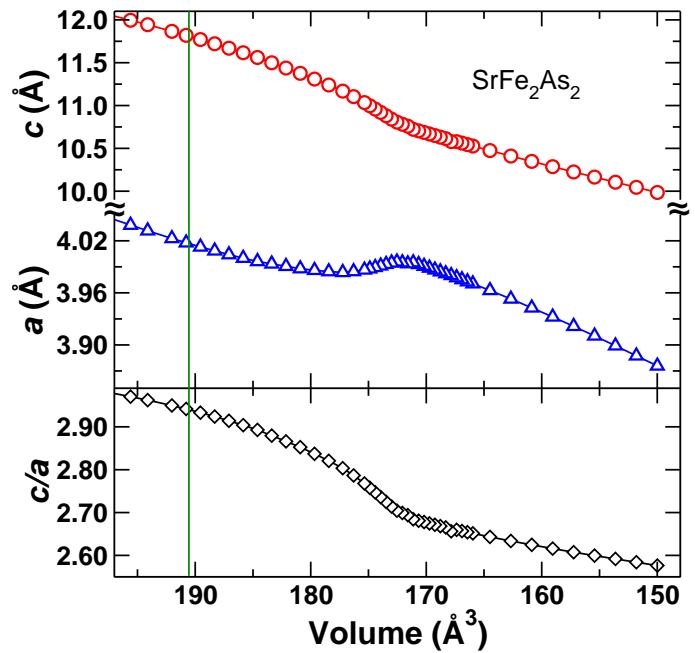}
\end{center}
\caption{\label{srfe2as2}(Color Online) Calculated $a$, $c$ and
$c/a$ as a function of unit cell volume for SrFe$_2$As$_2$
within LDA. At each volume, only the $c/a$ optimization has been
performed. The $c$ lattice parameter decreases throughout, while
the $a$ lattice parameter undergoes an anomalous expansion for
some reduced volumes before it continues to decrease again. The
vertical line denotes the experimental ambient condition volume.}
\end{figure}

\begin{figure}[t]
\begin{center}
\includegraphics[%
  clip,
  width=9cm,
  angle=-0]{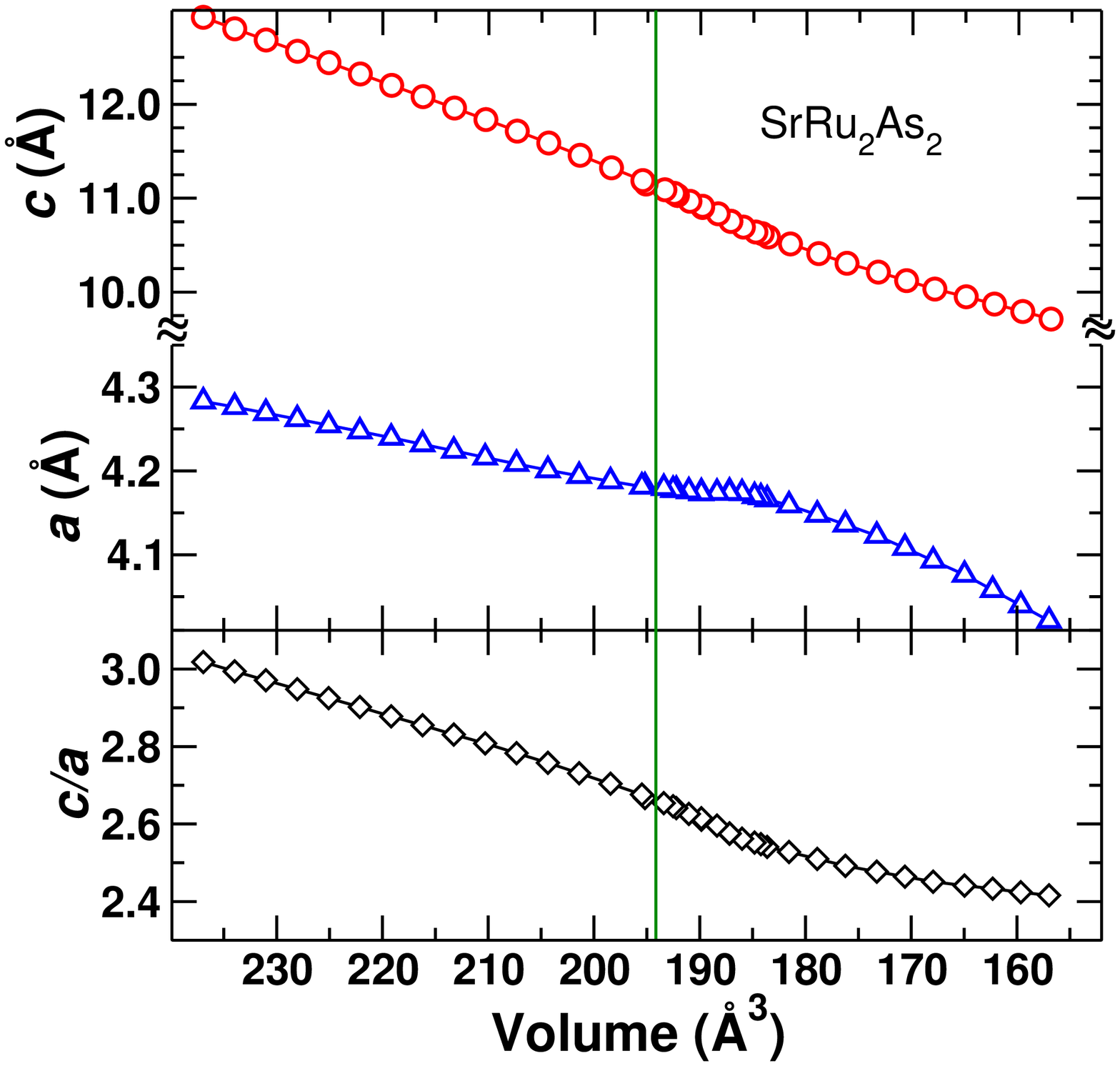}
\end{center}
\caption{\label{srru2as2}(Color Online) Calculated $a$, $c$ and
$c/a$ as a function of unit cell volume for SrRu$_2$As$_2$
within LDA. At each volume, only the $c/a$ optimization has been
performed. The vertical line denotes the experimental ambient condition
volume.}
\end{figure}

Collected in Fig.~\ref{exp} are the measured tetragonal lattice
parameters {\it a} and {\it c}, the axial ratio {\it c/a} and
the unit cell volume as a function of pressure for the parent
compound SrFe$_2$As$_2$ (the 10\% Ru-doped
SrFe$_{1.8}$Ru$_{0.2}$As$_2$ will be addressed later in
this report). For SrFe$_2$As$_2$, the lattice parameters {\it a} and {\it c} 
show an initial decrease with pressure  up
to 5 GPa. Upon further increase in pressure, anomalous
compression effects are observed with the lattice parameter $a$
expanding rapidly up to around 10 GPa, while the $c$ lattice
parameter continues to decrease. At this
juncture a strong decrease is witnessed in the {\it c} lattice parameter
to smaller values (a transition from a T$\longrightarrow$cT
phase), after which a normal compression behavior is observed up
to 20 GPa. The {\it c/a} ratio as a function of pressure shows
the onset of the structural phase transition around 10$\pm$1 GPa from a T-phase
(with {\it c/a} $\approx$ 2.85) to a cT-phase (with {\it c/a}
$\approx$ 2.65). Please refer to Table.\,1 and 2 in the supplementary
information for more details.

Similarly, an abrupt change in the interlayer As-As distance was
observed. 
However, this change of the interlayer distance can be 
acounted for only partially by the change of the lattice parameter $c$ (see
Fig.~\ref{exp}, lower panel).
This distance decreases from the rather large value
of 3.41\,{\AA} (at ambient pressure) to a value of 2.59\,{\AA} (at the
highest measured pressure of 19.6 GPa). Similar interatomic
distances are found for example in skutterudite CoAs$_{3}$ (2.49/2.56
\AA )\cite{Mandel1971} or filled skutterudite
LaFe$_{4}$As$_{12}$ (2.57/2.58 \AA ).\cite{Braun1980} 
These values are only slightly
larger than the single-bond distance of 2.52\,{\AA}  occurring in
$\alpha$-As.\cite{Donohue1974} The collapse of the lattice is in
turn, also accompanied by  a simultaneous pronounced decrease
of the Sr-As distances $d_{\rm SrAs}$ and Fe-As distances
$d_{\rm FeAs}$ by about 12\% and 4\%, respectively, as well as the
angles in the FeAs$_{4}$ tetrahedron (see
supplementary information). Electronic structure calculations
have shown that even minute shifts in $z$(As) and
thus the resulting Fe-As distances have strong impact on the
occupation of the Fe 3$d_{x^2-y^2}$ orbitals, and therefore, the
magnetic behavior in the $A$Fe$_2$As$_2$
phases.\cite{Krellner2008, Goldman2009} These effects should become
even more magnified during the observed phase transition. In
contrast, under applied external pressures up to 6 GPa, a
pronounced robustness of the Fe-As bonds  is recognized for
BaFe$_2$As$_2$\cite{Kimber2009} which crystallizes with a
significantly larger $c/a$ ratio of 3.35. However, the
Ba-As distances significantly contract (about 3\%).
Nevertheless, no indications of an onset of a lattice collapse
were detected for this compound.

The measured pressure-volume data were fit using a Birch-Murnaghan 
equation of state for the two separate phases (shown
as an inset in Fig.~\ref{exp}). The obtained equilibrium volume $V_{0}$,
bulk modulus $B_{\rm 0}$ and its pressure derivative $B'$ for
the tetragonal phase (0 - 8.6 GPa) are $V_{\rm 0}$ = 190.7 \AA$^{3}$, $B_{\rm 0}$ = 63.63
GPa and $B'$ = 2.51, respectively, and for the collapsed tetragonal
phase (11 - 19.5 GPa) are $V_{\rm 0}$ = 180.8 \AA$^{3}$, $B_{\rm 0}$ = 89.57 GPa and $B'$ = 2.51,
respectively. Our observation of the T$  \longrightarrow$ cT-phase
transition in SrFe$_2$As$_2$ along with the anisotropic
compressibility phenomena is akin to the reports on other 122
arsenides $AT_2$As$_2$ ($A$ = Ca, Ba, Eu; $T$ = Fe, Co).\cite{ca-pressure, ba-pressure1, ba-pressure2,
eu-pressure, co-pressure, Goldman2009}

\subsubsection{\rm Calculations: SrFe$_{2}$As$_{2}$}

Previous work on the tetragonal collapse in other members of the
$A$Fe$_2$As$_2$ family ($A$ = Ca, Ba, and Eu) have mostly
been experimental
studies.\cite{eu-pressure,ba-pressure1,ba-pressure2,co-pressure}
Analysis of the electronic structure of the Fe-As-based systems
using band structure calculations has been quite successful in
describing certain `general trends' like the tetragonal to
orthorhombic structural distortion, reduction in the magnetic
moment as a function of doping, etc., though DFT is less accurate in
reproducing certain details like the Fe-As bond
length.\cite{NJP} In 2009, Yildirim reported on results from
first-principles calculations for CaFe$_2$As$_2$ and noted
that the formation of the cT-phase is controlled by the Fe-spin
state.\cite{Yildirim} He suggested that pressure reduces the Fe-spin moment
which in turn weakens Fe-As bonding and strengthens As-As interaction
and therefore causes the collapse of the lattice parameters.
Here, also using first-principles calculations, we de-couple Fe
magnetism from the chemical bonding scenario and elucidate the
bond formation as the key feature.
To realize this, we calculated the change in the ground state
energy as a function of volume employing both LDA and GGA
and using the tetragonal symmetry without
invoking spin degrees of freedom explicitly. At each volume two kinds of optimization
were performed, one in which only the $c/a$ ratio was optimized
and another in which a simultaneous optimization of the $z$(As)
position was additionally carried out. Collected in Fig.~\ref{srfe2as2}
are the $a$ and $c$ lattice parameter
and the $c/a$ ratio as a function of volume obtained using the LDA with only $c/a$
optimization
(refer to Fig.\,1. in the supplementary information for details regarding the GGA calculations
and the effects of $z$(As) optimization).
The calculations reproduce the experimental results including the
transition from the ambient pressure T-phase to the cT-phase for
SrFe$_2$As$_2$ in accord with a dramatic jump in the axial
ratio $c/a$ from 2.85 $\rightarrow$ 2.65. Our calculations also
reproduce the observed anomalous expansion of the lattice
parameter  $a$ under pressure. This is the first theoretical
observation of the anomalous expansion phenomena for the
FeAs family of materials. By fitting the energy-volume ($E-V$) curves to
two separate Birch-Murnaghan equations of state (EOS) fits (not shown
here), we obtain a transition pressure of 11$\pm$1 GPa. 
The pressure ($P$) is obtained from the volume derivative of the EOS,
which is inverted to get $V(P)$.  Equating the enthalpies
$E[V(P)] + PV(P)$ of the two phases gives the transition pressure.
The
calculated value is in good agreement with the experimental value of 10$\pm$1
GPa. Calculations using GGA (refer to Fig.\,1. in supplementary information) with only $c/a$
optimization give essentially the same trend for $a$, $c$ and
$c/a$ as described above for LDA.

Relaxing additionally the $z$(As) position worsens the
quantitative description for both LDA and GGA. This behavior is
not surprising because, as mentioned previously, the Fe-As bond
length is quite sensitive and DFT fails in accurately
reproducing the experimental Fe-As bond length even for ambient
conditions.\cite{Mazin2008, Johannes2010} Another inadequacy of
DFT that is worth mentioning here is the equilibrium volume. It
is well known that LDA and GGA usually underestimate and
overestimate, respectively, the equilibrium volume by just a few
percent as compared to experiments. For SrFe$_2$As$_2$ both
LDA and GGA underestimate the equilibrium volume by 17\% and
10\%, respectively. This is quite unusual and at present there
exists no conclusive reasoning for such a behavior. One
possible explanation that is gaining more acceptance is the
presence of a nematic order, which is unaccessible using the
present-day DFT tools.\cite{nematic}

\subsubsection{\rm Experiment: SrFe$_{1.8}$Ru$_{0.2}$As$_{2}$}

Similar to hydrostatic external pressure, studies on the effects
of substitution on the transition metal Fe-site also show
anisotropic changes in the lattice parameters, with a
significant contraction of the $c$-axis length as compared to
$a$.\cite{SrFeRuAs,SrFeCoAs,Sefat2008,li2009} For example,
substitution of Fe with the isovalent, but larger Ru atom does
not introduce any additional charge into the system and is
suggested to simulate the effect of `chemical pressure',
alluding to the possibility of observing a T $\longrightarrow$ 
cT-phase transition in this substitution series. 
For the substitution series
SrFe$_{2-x}$Ru$_x$As$_2$ ($0 \leq x \leq 2$), a
significant but monotonous contraction of the lattice
parameter $c$ was reported, though no clear phase transition from a
T $\longrightarrow$ cT-phase was observed.\cite{SrFeRuAs} The lack
of a phase transition in SrFe$_{2-x}$Ru$_x$As$_2$  could be a
consequence of substitutional disorder in the samples and the
model of a chemically induced pressure volume effect is most
likely an oversimplified assumption.\cite{Rotter2010}
Nevertheless, we wanted to discern the idea of Ru substitution
acting as a `chemical pressure'. To this end, we have collected
the pressure dependence of the structural parameters $a$, $c$,
$c/a$ and unit cell volume for a 10\% Ru-doped sample
SrFe$_{1.8}$Ru$_{0.2}$As$_2$ (see Fig.~\ref{exp}). We have
chosen a rather small Ru content to keep the impurity/disorder
effects to a minimum. At room temperature and pressure,
SrFe$_{1.8}$Ru$_{0.2}$As$_2$ is in the T-phase with a $c/a
\approx$ 3.1. Similar to the parent compound, this sample is
also metallic and non-superconducting at ambient pressure.
Upon application of pressure, the system also
shows a transition to a cT-phase, though the transition pressure
is shifted downwards ($\approx$ 9$\pm$1 GPa) as compared
to SrFe$_2$As$_2$ (please refer to Table. 3 and 4 in supplementary information 
for more details). A fit to a Birch-Murnaghan equation of state 
was done for the tetragonal and the collapsed tetragonal phases.
The obtained equilibrium volume, bulk modulus $B_{\rm 0}$ and
its pressure derivative $B'$ for the tetragonal phase (0-7.3
GPa) are $V_{\rm 0}$ = 191.2 \AA$^{3}$, $B_{ \rm 0}$ = 63.3 GPa and $B'$ = 2.85, respectively, and
for the collapsed tetragonal phase (9-20 GPa) are $V_{\rm 0}$ = 183.1
\AA$^{3}$, $B_{\rm 0}$ = 83.6 GPa and $B'$ = 2.86, respectively. Comparing the
experimental results of SrFe$_2$As$_2$ and
SrFe$_{1.8}$Ru$_{0.2}$As$_2$, we can now perceive that the
10\% Ru substitution did indeed act as a `chemical pressure' in
the sample and reduced the external pressure needed for inducing
a T $\longrightarrow$ cT-phase transition in
SrFe$_{1.8}$Ru$_{0.2}$As$_2$ to 9 GPa as compared to the 10
GPa pressure needed for the (Ru-free) SrFe$_2$As$_2$ sample.

\subsubsection{\rm Calculations: SrRu$_{2}$As$_{2}$}

The bonding scenario for Ru in the end member of the
substitution series, SrRu$_2$As$_2$ is similar to that of Fe
in SrFe$_2$As$_2$. Moreover, SrRu$_2$As$_2$ is
non-magnetic at ambient conditions in contrast to the
paramagnetic SrFe$_2$As$_2$. Analyzing the effects of
external pressure in SrRu$_2$As$_2$ will therefore isolate
all other parameters and provide a purely chemical picture of
the transition. The $c/a$ ratio of SrFe$_2$As$_2$ is $\approx$ 3.15 at ambient
conditions, and is $\approx$ 2.65 for the cT
phase. On the contrary, SrRu$_2$As$_2$ has a ratio $c/a \approx$
2.68 at room temperature\cite{nath} and ambient pressure, and
therefore is already in the cT-phase at ambient conditions.
Therefore, the effect of external pressure on SrRu$_2$As$_2$
must be much weaker compared to SrFe$_2$As$_2$. Collected in
Fig.~\ref{srru2as2} are the changes in the lattice parameters
$a$, $c$ and $c/a$ obtained from LDA calculations as a function
of volume. Both $a$ and $c$ decrease rather monotonously with
the decrease in volume and do not show any intermediate
anamolous expansion of the $a$ lattice parameter. Moreover, in
contrast to the calculated results obtained above for
SrFe$_2$As$_2$ and in accordance to the generally accepted
trend, LDA slightly underestimates and GGA (Fig.\,2. in supplementary information)
slightly overestimates the equilibrium volume with respect to
the experiment.\cite{nath, SrFeRuAs}

Since this phase transition is not connected with a change in the symmetry, the nature of the transition 
can be first, second order, or even a continuous crossover. Moreover, the 
transition might be first order at $T$ = 0, but if the critical end point 
terminating the cT($P$) phase boundary line is below 300 K, one would 
only observe a continuous crossover at 300 K. The signatures expected for 
a first order transition, a second order one, and a crossover are respectively 
a jump, a jump in the derivative, and an S-shaped behavior in the relevant 
property, in the present case, the $c/a$ ratio. The data presented in 
Fig.~\ref{exp} do not evidence a sharp jump, instead they clearly suggest a 
jump in the derivative $d(c/a) /dT$, especially for pure SrFe$_{2}$As$_{2}$. 
Thus the present data indicate the transition in SrFe$_{2}$As$_{2}$ to be of 
second order type, instead of the well-established first order type in 
CaFe$_2$As$_2$.\cite{ca-pressure} This difference between CaFe$_2$As$_2$ 
and SrFe$_2$As$_2$ is in line with LDA calculations for the difference in 
the evolution of the total energy as a function of decreasing volume 
and $c/a$, where a second local minimum in $c/a$ is  much more pronounced 
in CaFe$_{2}$As$_{2}$ than in SrFe$_2$As$_2$.\cite{NJP}

\subsection{Chemical bonding}

Ternary compounds $AT_2X_2$ crystallizing with the tetragonal
ThCr$_2$Si$_2$ type of structure\cite{Ban1965} are
numerous\cite{Villars1997,Just1996} and have been the focus of
experimental and theoretical studies in solid state
sciences since several decades.\cite{Szytula1989} This research
originated with investigations on the Si and Ge-based family of
$AT_2X_2$ compounds,\cite{Ban1965, Bodak1965, Parthe1984} and
later embraced pnictogens as $X$ elements.\cite{Schlenger1971,
Mewis1977, Marchand1978, Jeitschko1980, Pfisterer1983} That
pnictogen-based members of this family of compounds are always
good for a surprise has already been recognized with the
discovery of superconducting LaRu$_2$P$_2$ ($T_{\rm c}$= 4.1 K) more than
two decades ago.\cite{Jeitschko1987} The crystal structure
comprises of $T_2X_2$ layers with edge-sharing $TX_{4}$
tetrahedra parallel to the $ab$ plane. These layers are
additionally separated by planes of metal atom $A$ (as can be
seen in Fig.\,\ref{elf}). 
Systematic structural and chemical observations
early on have resulted in classifying the A$T_2$$X_2$ compounds adopting
the ThCr$_2$Si$_2$ structure type into two branches: (i) One, with a
three-dimensional network built up of tetrahedral $TX_{4}$
layers, held together by $X-X$ bonds along the $c$ axis (two
apex $X$ atoms from 2 adjacent layers form the bond), with the
$A$ atoms embedded between these layers. This structural
peculiarity has been mainly observed for silicides and
germanides.\cite{Just1996} 
This branch is usually described as the ThCr$_2$Si$_2$ or CeGa$_2$Al$_2$ type.
(ii) The other branch exhibits a
rather two-dimensional layered structure with large separations
between the $X$ atoms along the $c$ axis. 
This branch is described as the TlCu$_2$Se$_2$ type. 
Interestingly, depending on the transition metal $T$, compounds containing
pnictogens (well documented for phosphides) have been found to
belong to both branches and to intermediate cases.
Variation and substitution of the transition metal $T$
(increasing the number of $d$ electrons) or the metal atom $A$
have been discovered as a means of  tuning this structural
peculiarity by causing a shrinkage of the $X-X$ distances and
bringing $X$ species in close bonding contacts. Furthermore, it
had also been discerned that the geometrical constraints due to
the changes in the sizes of the constituent atoms alone was not
adequate to facilitate the $X-X$ bond formation. These analyses
early on evoked questions about the electronic structure
governing the underlying bonding
situations\cite{Mewis1980,Jeitschko1980} and the observed
physical properties.\cite{Jeitschko1987b, Pfisterer1983,
Reehuis1998} Absence of pnictogen-pnictogen bonds indicated by
large $c/a$ rations leads to a 
composite-like structural arrangement of the [$T_2X_2$] layers held
together by more or less ionic interactions mediated by the
electropositive $A$ cations. Members of the TlCu$_2$Se$_2$ branch could
therefore be qualitatively described within the Zintl
concept\cite{Schaefer1973} by assigning a formal oxidation number
-3 to the most electronegative element (pnictogen) resulting in
a polyanion [$T_2X_2$]$^{2-}$ counterbalanced by a cation $A^{2+}$
(e.g. Sr$^{2+}$Fe$^{2+}$Fe$^{2+}$P$^{3-}$P$^{3-}$). In contrast,
compounds of the TlCu$_2$Se$_2$ family
exhibiting pnictogen-pnictogen bonds are
preferentially formed by the late transition metals. The
corresponding qualitative view in light of the Zintl concept is
now based on diatomic [$X_2$]$^{4-}$ entities (e.g.
Ca$^{2+}$Ni$^{1+}$Ni$^{1+}$[P$_2$]$^{4-}$ as discussed in
Refs.\,\onlinecite{Jeitschko1987,Jeitschko1987b}). However,
regarding the $T-T$ bonding interactions a slightly different
trend was inferred from atomic distances and occupation of
simple and idealized molecular orbital schemes, suggesting a
more pronounced covalency for Fe-Fe bonds than for Ni-Ni bonds
(Refs.\,\onlinecite{Jeitschko1984,Jeitschko1985,Jeitschko1985b}).
In a similar way of reasoning the chemical bonding was
qualitatively rationalized in As-based
compounds.\cite{Jeitschko1987b,Pfisterer1983} Hoffmann
highlighted this genuine occurrence of making and breaking of a
diatomic ($X-X$) bond in the solid state for the A$T_2$$X_2$
structures, and pioneered the
investigation of the chemical bonding situation in phosphides
based on extended H\"uckel calculations, discovering 
structural flexibility due to a subtle interplay between
pnictogen-pnictogen and pnictogen-metal bonding with the
additional ingredient of packing requirements of the large $A$
atoms.\cite{Hoffmann1985}

\subsubsection{Chemical bonding in $k$-space}
\begin{figure}[b]
\begin{centering}
\includegraphics[clip,width=9cm]{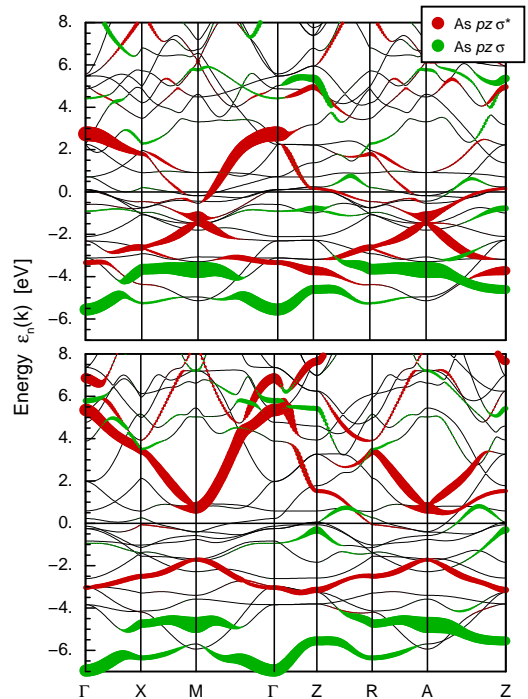}
\par\end{centering}
\caption{\label{fig:fatbands_B_AB}(Color online) Bonding
($\sigma$, green or light gray) and antibonding ($\sigma^{*}$
red or dark gray) As 4$p_{z}$ {}``fat bands'' for
SrFe$_2$As$_2$. The Fermi level is at $0$\,eV. Upper panel:
T-phase. Lower panel: cT-phase.}
\end{figure}

In order to investigate the development of an interlayer
chemical bond between As $p_{z}$ orbitals in SrFe$_2$As$_2$
as we go from the ambient pressure T-phase to the cT-phase, we
analyze the orbital character of the band structure ({}``fat
bands''). Usually, fat bands are obtained by suitable projection
of the extended wave function onto test orbitals with a certain
character, which in a local-orbital code naturally are chosen to
be the local orbitals themselves. Atom-centered projectors do not
probe the phase relations between different sites, though. In
order to obtain this phase information, which contains the
chemical interpretation in terms of bonding characteristics,
projectors are needed, which contain several sites. We choose
projectors, which include As 4$p_{z}$ orbitals of a pair of As
atoms sitting on top of each other across the Sr interlayer
spacer. The two combinations are bonding
$\Phi_{\sigma}=\frac{1}{\sqrt{2}}\left(\Phi_{1}-\Phi_2\right)$
and antibonding
$\Phi_{\sigma^{*}}=\frac{1}{\sqrt{2}}\left(\Phi_{1}+\Phi_2\right)$
within the pair. Here, the negative sign gives bonding because of the
odd parity of the $p$-orbitals. The corresponding fat bands
probe the phase correlation between the two As $p_{z}$ orbitals
but do not probe the phase between pairs of As atoms within the
layer. Due to the unitarity of the transformation from orbitals
to projectors the resulting bonding and anti-bonding orbital
weights sum up to the {}``standard'' orbital-projected weights,
which opens not only the possibility to analyse the band
structure but also the projected density of states according to
bonding characteristics. 

Fig.\,\ref{fig:fatbands_B_AB} shows the
$\sigma$ and $\sigma^{*}$ fat bands for the ambient pressure
T-phase and for the high pressure cT-phase of
SrFe$_2$As$_2$. In both phases the bonding part of the
4$p_{z}$ bands (green or light grey) are mostly occupied. In the
T-phase the antibonding bands are significantly occupied, while
they get pushed up close to and above the Fermi level in the
cT-phase. There is one antibonding band remaining below the
Fermi level, however the total weight of occupied antibonding
bands is strongly reduced for the cT-phase as compared to the
T-phase.

\begin{figure}[t]
\noindent \begin{centering}
\includegraphics[width=8cm]{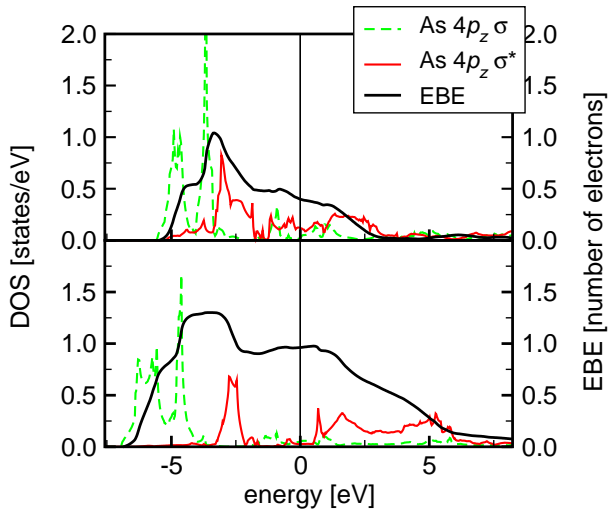}
\par\end{centering}
\caption{\label{fig:bdos}(Color online) The $\sigma$ (dashed,
green color or light grey) and $\sigma^{*}$ (solid, red color or
dark grey) density of states for the T phase (upper panel) and
for the cT-phase (lower panel) of SrFe$_2$As$_2$. The solid
(black) line shows the number of excess bonding electrons, EBE
(see text for explanation).}
\end{figure}

\begin{figure}[t]
\begin{centering}
\includegraphics[clip,width=8.5cm]{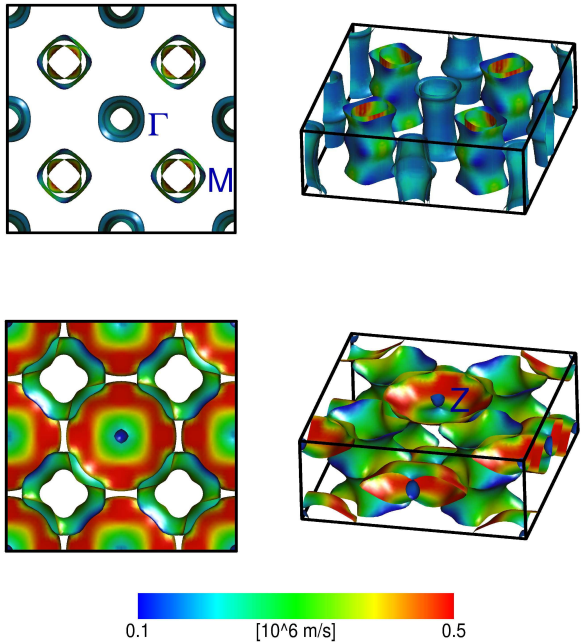}
\par\end{centering}
\caption{\label{fig:FS}(Color online) Fermi surfaces for the
tetragonal T (upper panel) and collapsed tetragonal cT (lower panel)
phase. The color code shows the magnitude of the Fermi
velocity.}
\end{figure}

In order to quantify the bond strength we plot the difference
of the integrated $\sigma$ and $\sigma^{*}$ projected density of
states (DOS) along with the projected DOS in
Fig.\,\ref{fig:bdos}. In an idealized case of completely
occupied bonding and completely unoccupied antibonding states
this difference of energy-resolved occupations will be
monotonously increasing up to a value of two (spin degeneracy)
at the chemical potential, signifying that two electrons form
the bond, and will monotonously decrease with increasing energy
away from the Fermi level, to reach zero if all bands are filled
with a total number of four electrons. The value of this
difference occupation at the Fermi level is an estimate of
chemical binding in terms of excess bonding electrons (EBE).
Fig.\,\ref{fig:bdos} shows that the number of EBE
is only 0.4 in the T-phase but is reaching nearly 1 in
the collapsed phase. It turns out that the number of EBE
increases by $\approx0.7$ immediately before the collapse
occurs, showing that the driving force of the collapse is the
onset of 4$p_{z}$ interlayer bond formation.

Returning to the band structure, besides the increased band width
in the cT-phase compared to T-phase, there are additional
changes occuring at the Fermi level. In the T-phase there is a
strongly antibonding band starting at -0.5\,eV at the M-point
going to 3\,eV at the $\Gamma$-point. This band is hybridized with
the Fe 3$d_{x^{2}-y^{2}}$ orbitals. In the cT-phase this
particular band now begins at around 1eV above the Fermi level
and has a larger dispersion going up to 7\,eV. Most importantly,
this band that formed a Fermi surface in the T-phase, gets
removed from the Fermi level for the cT-phase, which results in
a change of the Fermi surface topology (Fig.\,\ref{fig:FS}). In
general the Fe bands do not change as drastically as the As
bands across the transition. Most notably the Fe band width is
not following the strong increase of the As band width. From a
chemical point of view, the Fe layer stays inactive in the
collapse. However, from a physical point of view, the Fe bands,
especially the Fermi surface, changes from a 2D to a 3D system.
Of the two cylindrical sheets around the M-point (T-phase) only
one remains after the collapse, while of the three cylinders
around $\Gamma$ one vanishes, the second becomes a very small
pocket, and the third forms a large pocket around the Z-point.
This large pocket shows a quite 3-dimensional distribution of
Fermi velocities. The plasma frequencies change from
$\Omega_{xy}=2.81$\,eV and $\Omega_{z}=0.96$\,eV in the T-phase to
$\Omega_{xy}=2.56$\,eV and $\Omega_{z}=3.81$\,eV in the cT-phase. In
fact the ratio $\lambda$ between in-plane and out-of-plane
plasma frequencies goes from $\lambda\approx3$ to
$\lambda\approx0.7$. Interestingly, the individual bands behave
rather differently. The still more cylindrical band around M
has a ratio $\lambda\approx1.2$ while the pocket around  Z has
$\lambda\approx0.48$, which makes this band actually tend to be
more one-dimensional along the $z$ axis, which is another aspect
in the formation of interlayer As-As bonds.

\subsubsection{Chemical bonding in real space}

\begin{figure}[t]
\begin{center}
\includegraphics[%
  clip,
  width=9cm,
  angle=-0]{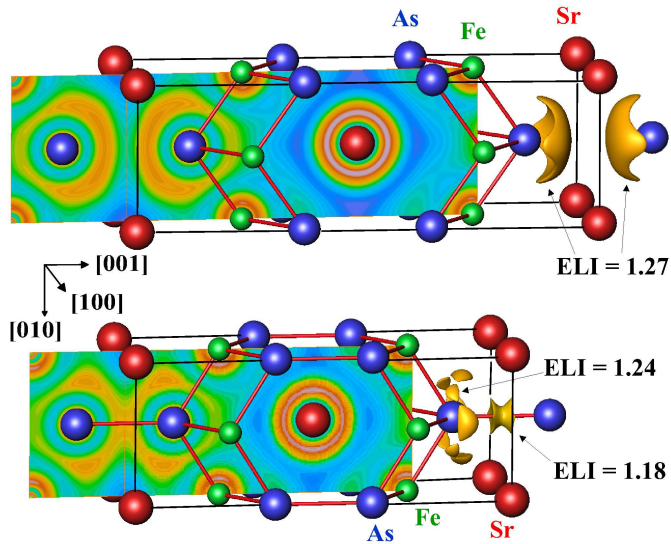}
\end{center}
\caption{\label{elf}(Color online) ELI-D of the ambient pressure
T-phase (top panel) and the cT-phase at high pressure (bottom
panel) of SrFe$_2$As$_2$. The slices in the left part reveal the
distribution of ELI-D between the As atoms. The absence of As-As
bonding in the T-phase is illustrated by the isosurfaces of
$\Upsilon$=1.27, while the formation of As-As bonds in the
cT-phase is illustrated by the isosurfaces of $\Upsilon$=1.18
.}
\end{figure}

The electron localizability indicator (ELI,
$\Upsilon$) was evaluated in the ELI-D representation
according to Refs.\,\onlinecite{Kohout04,eli3} with an ELI-D module.
The ELI-D distribution for the ambient condition T-phase shown in the upper panel of
Fig.~\ref{elf} has four distinct features. The valence (5th)
shell of the Sr atoms is absent suggesting the formation of
the Sr cation and the transfer of these electrons to the
[Fe$_2$As$_2$] anion. The penultimate (fourth) shell of Sr is
not specially structured\cite{eli4,eli5} indicating that the
electrons of this shell do not participate in the bonding
interactions in the valence region. Between the As atoms, two
distinct maxima of ELI-D are observed illustrating the absence
of As-As bonds and the non-bonding (lone-pair-like) interaction
between the neighboring [Fe$_2$As$_2$] anions. The
structuring of the penultimate (third) shell of the Fe atoms
towards the closest As is the fingerprint of the Fe-As bonding
within this anion. For the high-pressure cT-phase (lower panel
of Fig.\,\ref{elf}) in addition to the Coulomb interaction
between the Sr cations and the [Fe$_2$As$_2$] anions, there
is also bonding by the electrons of the penultimate (fourth) Sr
shell (cf.\,structuring of this shell in Fig.~\ref{elf}, lower panel). 
This observation is similar to the one
for Eu in EuRh$_2$Ga$_{8}$.\cite{eli6}. The distinct
maxima found close to the Fe-As contacts (ELI-D isosurface with
$\Upsilon$=1.24) shows the formation of covalent bonds within
the [Fe$_2$As$_2$] anion. The formation of the As-As bonds
between the neighboring anions is visualized in real space by a
concentration of ELI-D close to the middle point of the As-As
contact (ELI-D isosurface with $\Upsilon$=1.18), which is very
similar in topology to the ELI-D distribution in the simple example
of the F$_2$
molecule.\cite{eli7,eli8}

\section{Summary}

In conclusion, we have studied the transition under high pressure
from a tetragonal (T) to a collapsed tetragonal (cT) phase in
SrFe$_2$As$_2$ and SrFe$_{1.8}$Ru$_{0.2}$As$_2$, using
diamond anvil cells and powder XRD measurements up
to 20 GPa at room temperature. We observe an isostructural phase
transition from a T-phase to a cT-phase at 10 GPa for
SrFe$_2$As$_2$ and at 6 GPa for
SrFe$_{1.8}$Ru$_{0.2}$As$_2$. Both materials show an
anomalous compression behavior (anisotropic changes in the unit
cell dimensions) under pressure with the lattice parameter $c$ 
decreasing continuously while the lattice parameter $a$ 
increases for a certain pressure range. Our observation is akin
to previous reports\cite{ca-pressure, ba-pressure1,
ba-pressure2, eu-pressure, co-pressure} on other 122 systems
$AT_2$As$_2$ ($A$ = Ca, Ba, Eu; $T$ = Fe, Co). From our
experiments, we note that Ru substitution of the Fe-site works
as `chemical pressure' thereby reducing the amount of external
pressure needed to obtain the T $\longrightarrow$ cT-phase
transition in SrFe$_{1.8}$Ru$_{0.2}$As$_2$. Band structure
calculations reproduce the isostructural phase transition
observed in experiments, including the anisotropic changes in
the unit cell dimensions. Detailed analyses in $k$-space (Fermi
surfaces, band structure and density of states) and as well as in
real space (ELI-D) of the bonding scenario within and between
the FeAs layers in both the T and cT-phases presently provides a
comprehendible picture of the driving force behind the observed
lattice collapse. The actual phase transition can be
rationalized in the following way. Since the
covalently bonded rather rigid transition metal-pnictogen layers
are separated by the large cations (e.g. Sr$^{2+}$), short As-As
bonds along the $c$ axis cannot be formed due to geometric
reasons in the ambient pressure T-phase. The predominantly ionic
interaction between the cations and the polyanion makes these
layered compounds soft along the stacking $c$ axis. Thus,
applying pressure leads to a decrease of the distance between
the layers and therefore the As-As separation. When a critical
As-As distance is reached to enable a sufficiently high enough
orbital overlap of the 4$p_{z}$ orbitals, promptly the bonding
interactions dominate and the phase transition occurs. Similar
to  the case  previously studied for phosphorus-based
compounds,\cite{Johrendt1997,Huhnt1997} electronic structure
calculations outline a scenario where a stabilization of the
As-As bonding states stabilize Fe-As antibonding states which
have lifted up the former ones to energies near the Fermi level.
Consequently, the Fe-As bonds become weaker and the lattice
parameter $a$ increases whereas $c$ decreases at the phase
transition.

\section*{Acknowledgement}
We acknowledge stimulating discussions with W. Schnelle. 
This research has been partly funded by DFG within SPP
1458. We thank H. Borrmann and U. Burkhardt for performing powder
X-ray diffraction and metallographic characterization as well as
EPMA of the samples used in this study.

\bibliographystyle{apsrev}
\bibliography{paper}

\end{document}